\documentclass[superscriptaddress,aps,amsmath,amssymb,showpacs,showkeys]{revtex4}
\usepackage[dvips]{graphicx,color}
\usepackage[utf8]{inputenc}
\usepackage{multirow}
\usepackage{times}
\usepackage{xcolor}
\usepackage[%
  colorlinks=true,
  urlcolor=blue,
  linkcolor=red,
  citecolor=blue
]{hyperref}
\usepackage{orcidlink}

\begin{document}

\title{Violation of Hod's Conjecture and Probing it with Optical properties of a $5$-D black hole in Einstein Gauss-Bonnet Bumblebee theory of gravity}

\author{Dhruba Jyoti Gogoi\orcidlink{0000-0002-4776-8506}}
\email[Email: ]{moloydhruba@yahoo.in}

\affiliation{Department of Physics, Moran College, Moranhat, Charaideo 785670, Assam, India.}
\affiliation{Theoretical Physics Division, Centre for Atmospheric Studies, Dibrugarh University, Dibrugarh
786004, Assam, India.}

\begin{abstract}
In this work, the quasinormal modes of a $5$-D black hole in Einstein Gauss-Bonnet Bumblebee theory of gravity have been investigated with the help of the Pad\'e averaged higher order WKB approximation method and the validity of Hod's conjecture has been studied. It is found that the presence of Lorentz symmetry breaking due to the Bumblebee field favours Hod's conjecture. But in the case of the Gauss-Bonnet term, an increase in the coupling parameter increases the chances of violation of Hod's conjecture. We further investigated the optical properties of the black hole {\it viz.}, shadow and emission rate. It is found that a black hole with a lower lifetime favours Hod's conjecture.
\end{abstract}


\maketitle
\section{Introduction} \label{sec01}

For centuries, physicists have delved into the mysteries of the Universe's fundamental nature. The landmark discovery of Gravitational Waves (GWs) in 2015 profoundly validated General Relativity (GR), the cornerstone theory of classical gravity. While GR has admirably explained celestial phenomena and compact objects such as black holes, its inability to address gravity at the quantum level, where it lacks renormalizability at the UV scale, highlights its limitations. In contrast, the Standard Model (SM) of particle physics successfully describes fundamental particles and their interactions at the quantum level. The ambition to unify these two successful yet incompatible theories has given rise to Quantum Gravity (QG) theories, which can only be directly probed at the Planck scale ($\approx 10^{19}$ GeV). Although direct experimental validation of QG remains elusive, certain effects, like the potential breaking of Lorentz symmetry \cite{kost2004}, might manifest at lower energy scales \cite{casana2018}. Additionally, observational data in the infra-red scales have hinted at discrepancies with GR, suggesting a need for modifications in that regime as well. This article explores the challenges and prospects of integrating Gauss-Bonnet extended GR with the SM within the framework of QG theories, offering deeper insights into the fundamental mechanisms governing the Universe. Specifically, our focus lies on scalar perturbations within the 5D black hole spacetime and their optical characteristics in this unified extended theory.

Such theories have been explored in various contexts previously. The idea of Lorentz symmetry violation has been investigated in several theories like loop quantum gravity, noncommutative field theories, string theory, and the standard-model extensions (SME). The SME is a broad field theoretical framework that includes both the SM and GR, containing terms in the Lagrangian indicating possible Lorentz symmetry violations. In the gravitational aspect of the SME, prior studies have examined the implications of Lorentz symmetry breaking, as seen in several researches such as Ref.s \cite{bluhm2005,bailey2006, bailey2009,tso2011,kost2009,maluf2013,maluf2014, santos2015}. Similarly, the effects of Lorentz violation on GWs also have been investigated \cite{kost2016, kost2016_2}. The simplest extended gravitational field theories involving the spontaneous breaking of Lorentz and diffeomorphism symmetries are known as Bumblebee models \cite{kost2004}. These models include a single vector field, known as the Bumblebee field, with a non-zero vacuum expectation value that breaks Lorentz symmetry. Recent studies have investigated spherically symmetric vacuum solutions to the Einstein field equations in the presence of spontaneous Lorentz symmetry breaking due to the Bumblebee field \cite{casana2018}. These studies also examined classical tests, showing that Lorentz violation introduces corrections even without a massive gravitational source and deforms the background spacetime. Another study examined particle motion in Snyder noncommutative spacetime structures under Lorentz violation \cite{kumar2021}. Additional references related to Lorentz symmetry violation can be found in \cite{kanzi2019,gullu,jha2022,Jusufi19}.

It is common knowledge that M/string theory is one of the most promising hypotheses for unifying all interactions \cite{Pavluchenko:2016wvi}. At lower energy levels, super-symmetric string theory generates effective field theories that correct Einstein's GR with quadratic and higher curvature terms \cite{wiltshire}. These corrections are generalized in Lovelock's theory \cite{lovelock}, where GR is just the first-order approximation that cannot account for strong gravity effects, such as those occurring in the interior of a black hole. The only combination of quadratic terms that results in a ghost-free, nontrivial gravitational interaction is the Gauss-Bonnet invariant $\mathcal{G}$ \cite{zwiebach}, which represents the second-order approximation of the gravity theory. The parameter $\alpha$ preceding $\mathcal{G}$ can be regarded as an expansion parameter. However, since this invariant is a topological invariant in four dimensions, it does not contribute to the gravitational dynamics in a four-dimensional spacetime with $D=4$.

Glavan and Lin recently introduced a groundbreaking four-dimensional Einstein-Gauss-Bonnet gravity theory using a regularization scheme \cite{lin2020}. Their approach involves rescaling the expansion parameter $\alpha$ by $\alpha/(D-4)$ in $D$ dimensions and defining the four-dimensional theory at the field equation level, in the limit as $D$ approaches 4. However, some scholars have pointed out drawbacks to this regularization scheme \cite{shu,arrechea}. Nonetheless, other scholars have suggested various remedies to overcome these objections \cite{hennigar} and have successfully obtained the same black hole metric as the original version \cite{lin2020}, thus validating it as a viable solution.

In their work \cite{fernandes}, Fernandes derived the Reissner-Nordstr\"{o}m black hole solution in the presence of EGB gravity coupled to a Maxwell electric field. Similarly, Yang {\it et al.} \cite{yang} obtained the Born-Infeld black holes in EGB gravity that is minimally coupled to the Born-Infeld nonlinear electromagnetic field. Feng {\it et al} \cite{feng}, on the other hand, investigated the cosmological solution with EGB gravity coupled to a scalar field. In Refs. \cite{kumar2004}, Bardeen-like and Hayward-like black holes were derived. Additionally, Wei {\it et al} \cite{wei2021} studied rotating black hole solutions, while some other types of black hole solutions were explored in Refs. \cite{ghosh}.

 In 2020, an exact Kerr-like solution was found by solving the Einstein-bumblebee gravitational field equations and its black hole shadow was studied explicitly \cite{ding2020}. Li and \"{O}vg\"{u}n \cite{li} investigated the weak gravitational deflection angle of relativistic massive particles by this Kerr-like black hole. Furthermore, Jha and Rahaman \cite{jha} extended this Kerr-like solution to the Kerr-Sen case.

This paper focuses on studying the effective field theory of Einstein-Gauss-Bonnet gravity coupled with bumblebee fields to investigate the suppressed effects that emerge from the underlying unified quantum gravity theory at low energy scales. Here, we will investigate the quasinormal modes, shadow and emission rates of the black hole in the theory in $5$--D. We have implemented the Pad\'e averaged WKB approximation method up to sixth order to calculate the quasinormal modes, which is known to provide a good approximation, but in some cases, {\it e.g.}, when overtone number is higher than the multipole moment, it may not be adequate to accurately calculate quasinormal modes \cite{Daghigh2012}.  The quasinormal modes refer to complex numbers that are linked to the emission of gravitational waves from perturbed, massive objects in the Universe \cite{Vishveshwara, Press, Chandrasekhar_qnms}. The real component of the quasinormal modes is related to the emission frequency, whereas the imaginary component is related to its damping. Different studies have explored the properties of quasinormal modes and gravitational waves in black holes that exist in various modified gravity theories \cite{Ma, Pedrotti:2024znu, gogoi1, gogoi2, Liang_2017, qnm_bumblebee, gogoi3, Graca, Zhang2, lopez2020, Liang2018, Hu, hemawati2022, gogoi4}. In Ref. \cite{qnm_bumblebee}, the impact of Lorentz violation on quasinormal frequencies has been investigated for black holes in Bumblebee gravity. In another work \cite{Gogoi:2022wyv}, quasinormal modes and Hawking radiation sparsity of de Sitter and anti-de Sitter black hole solutions have been investigated in the presence of the Generalised Uncertainty Principle and topological defects in Bumblebee gravity.

The work is organised as follows. In Section \ref{sec02}, we discuss Einstein-Gauss-Bonnet-bumblebee theory in brief. In Section \ref{sec03}, we discuss the black hole solution in this theory in 5D. In Section \ref{sec04}, we discuss the scalar perturbation, quasinormal modes and validity of Hod's conjecture. Optical properties of the black hole and its relation with Hod's conjecture are discussed in Section \ref{sec05}. Finally in Section \ref{section6}, we conclude the work with a brief overview of the results and possible future prospects.
Throughout the paper we have 
considered $ \kappa=G=c=1$.

\section{Einstein-Gauss-Bonnet-Bumblebee theory: A quick review} \label{sec02}

Given that EGB gravity is a second-order approximation of the theory of gravity and can effectively account for strong gravity effects, it is natural to consider the coupling of SM to EGB gravity to explore possible LV effects in the strong gravity sector. The Bumblebee gravity model introduces a Bumblebee vector field $B_{\mu}$ with a non-zero vacuum expectation value, which triggers spontaneous Lorentz symmetry breaking in the gravitational sector through a specific potential. In $D-$dimensional spacetime, the action of Einstein-Gauss-Bonnet gravity coupled to this Bumblebee field is expressed as \cite{Ding2022},
\begin{eqnarray}
\mathcal{S}=
\int d^Dx\sqrt{-g}\Big[\frac{R}{2 }+\frac{2\alpha}{D-4} \mathcal{G}+\frac{\varrho}{2 } B^{\mu}B^{\nu}R_{\mu\nu}-\frac{1}{4}B^{\mu\nu}B_{\mu\nu}
-V(B_\mu B^{\mu}\mp b^2)+\mathcal{L}_M\Big], \label{action}
\end{eqnarray}
Here, $R$ represents the Ricci scalar and $\alpha$ is an expansion parameter. 

The Gauss-Bonnet invariant $\mathcal{G}$ is given by,
\begin{eqnarray}\label{}
\mathcal{G}=R_{\mu\nu\tau\sigma}R^{\mu\nu\tau\sigma}-4R_{\mu\nu}R^{\mu\nu}+R^2,
\end{eqnarray}
which represents the second-order approximation of the theory of gravity and may effectively account for strong gravity effects. In four dimensions, it is a topological invariant and does not contribute to gravitational dynamics. However, Glavan and Lin \cite{lin2020} showed that if one uses a regularization scheme to rescale $\alpha$ by $\alpha/(D-4)$ and take the limit $D\rightarrow4$, this Gauss-Bonnet term can have a non-trivial impact.

The coupling constant $\varrho$ plays a significant role in the non-minimal gravity interaction with the Bumblebee field $B_\mu$. The term $\mathcal{L}_M$ encompasses any possible interactions with matter or external currents.
The positive real constant, denoted as $b$, plays a significant role in triggering Lorentz violation by means of the potential $V(B_\mu B^{\mu}\mp b^2)$, which is responsible for imparting a non-zero vacuum expectation value (VEV) to the Bumblebee field $B_{\mu}$ and breaking Lorentz symmetry spontaneously. The potential attains its minimum value when $B^{\mu}B_{\mu}\pm b^2=0$ and $V'(b_{\mu}b^{\mu})=0$, which ensures the destruction of the $U(1)$ symmetry. At this point, the field $B_{\mu}$ acquires a non-zero VEV, $\langle B^{\mu}\rangle= b^{\mu}$. Furthermore, $b^{\mu}$ is another vector that varies with the spacetime coordinates and possesses a constant value of $b_{\mu}b^{\mu}=\mp b^2$, where the $\pm$ signs signify that $b^{\mu}$ is either timelike or spacelike, respectively.
The Bumblebee field has a strength described by the equation 
\begin{eqnarray}
B_{\mu\nu}=\partial_{\mu}B_{\nu}-\partial_{\nu}B_{\mu}.
\end{eqnarray}
The anti-symmetry of $B_{\mu\nu}$ leads to the constraint  \cite{bluhm}
\begin{eqnarray}
\nabla ^\mu\nabla^\nu B_{\mu\nu}=0,
\end{eqnarray} 
which implies that conservation law remains valid even though the potential $V$ breaks the $U(1)$ gauge symmetry. In both Minkowski and Riemann spacetime, this model provides a dynamic theory which generates a photon as a Nambu-Goldstone boson for spontaneous Lorentz violation, without introducing any ghosts \cite{bluhm2005}.
Varying the action (\ref{action}) with respect to the metric yields the gravitational field equations
\begin{eqnarray}\label{einstein0}
G_{\mu\nu}=R_{\mu\nu}-\frac{1}{2}g_{\mu\nu}R=  T_{\mu\nu}^B+2\alpha  T^{GB}_{\mu\nu}+  T_{\mu\nu}^M,
\end{eqnarray}
where the Bumblebee energy-momentum tensor $T_{\mu\nu}^B$ is
\begin{eqnarray}\label{momentum}
&&T_{\mu\nu}^B=B_{\mu\alpha}B^{\alpha}_{\;\nu}-\frac{1}{4}g_{\mu\nu} B^{\alpha\beta}B_{\alpha\beta}- g_{\mu\nu}V+
2B_{\mu}B_{\nu}V'\nonumber\\
&&+\varrho\Big[\frac{1}{2}g_{\mu\nu}B^{\alpha}B^{\beta}R_{\alpha\beta}
-B_{\mu}B^{\alpha}R_{\alpha\nu}-B_{\nu}B^{\alpha}R_{\alpha\mu}\nonumber\\
&&+\frac{1}{2}\nabla_{\alpha}\nabla_{\mu}(B^{\alpha}B_{\nu})
+\frac{1}{2}\nabla_{\alpha}\nabla_{\nu}(B^{\alpha}B_{\mu})
-\frac{1}{2}\nabla^2(B^{\mu}B_{\nu})-\frac{1}{2}
g_{\mu\nu}\nabla_{\alpha}\nabla_{\beta}(B^{\alpha}B^{\beta})\Big],
\end{eqnarray}
and the Gauss-Bonnet energy-momentum tensor $T^{GB}_{\mu\nu}$ is,
\begin{eqnarray}\label{momentumG}
&&T_{\mu\nu}^{GB}=4R_{\alpha\beta}R^{\alpha\;\beta}_{\;\mu\;\nu}-2R_{\mu\alpha\beta\gamma}
R_{\nu}^{\;\alpha\beta\gamma}+4R_{\mu\alpha}R^{\alpha}_{\;\nu}-2RR_{\mu\nu}
+\frac{1}{2}g_{\mu\nu}\mathcal{G}.
\end{eqnarray}
The prime denotes differentiation with respect to the argument,
\begin{eqnarray}
V'=\frac{\partial V(x)}{\partial x}\Big|_{x=B^{\mu}B_{\mu}\pm b^2}.
\end{eqnarray}
Varying instead with respect to the Bumblebee field generates the Bumblebee equations of motion (supposing that there is no coupling between the Bumblebee field and $\mathcal{L}_M$),
\begin{eqnarray}\label{motion}
\nabla ^{\mu}B_{\mu\nu}=2V'B_\nu-\varrho B^{\mu}R_{\mu\nu}.
\end{eqnarray}

The contracted Bianchi identities ($\nabla ^\mu G_{\mu\nu}=0$) lead to conservation of the total energy-momentum tensor
\begin{eqnarray}\label{}
\nabla ^\mu T_{\mu\nu}=\nabla ^\mu\big( T^B_{\mu\nu}+2\alpha T^{GB}_{\mu\nu}+T^M_{\mu\nu}\big)=0.
\end{eqnarray}

In the next sections, we derive the black hole solution and cosmological solution by solving gravitational equations in this Einstein-Gauss-Bonnet-Bumblebee model.

\section{Black hole solution in Einstein-Gauss-Bonnet-Bumblebee gravity} \label{sec03}
In this section, we suppose that there is no matter field and the Bumblebee field is frosted at its VEV like in Refs \cite{casana2018, Bertolami2005}, i.e., it is
\begin{eqnarray}
B_\mu=b_\mu,
\end{eqnarray}
then the specific form of the potential controlling its dynamics is irrelevant.
And as a result, we have $V=0,\;V'=0$. Then the first two terms in Eq. (\ref{momentum}) are like those of the electromagnetic field, the only distinctive are the coupling items to Ricci tensor. Under this condition,  Eq. (\ref{einstein0}) leads to gravitational field equations \cite{Ding2022}
\begin{eqnarray}\label{bar}
G_{\mu\nu}=2\alpha  T^{GB}_{\mu\nu}+  (b_{\mu\alpha}b^{\alpha}_{\;\nu}-\frac{1}{4}g_{\mu\nu} b^{\alpha\beta}b_{\alpha\beta})+\varrho\Big(\frac{1}{2}
g_{\mu\nu}b^{\alpha}b^{\beta}R_{\alpha\beta}- b_{\mu}b^{\alpha}R_{\alpha\nu}
-b_{\nu}b^{\alpha}R_{\alpha\mu}\Big)
+\mathcal{B}_{\mu\nu},
\end{eqnarray}
with
\begin{eqnarray}\label{barb}
&&\mathcal{B}_{\mu\nu}=\frac{\varrho}{2}\Big[
\nabla_{\alpha}\nabla_{\mu}(b^{\alpha}b_{\nu})
+\nabla_{\alpha}\nabla_{\nu}(b^{\alpha}b_{\mu})
-\nabla^2(b_{\mu}b_{\nu})-g_{\mu\nu}\nabla_\alpha\nabla_\beta(b^\alpha b^\beta)\Big].
\end{eqnarray}
The static spherically symmetric black hole metric in a $5$ dimensional spacetime has the general form
\begin{eqnarray}\label{metric}
&&ds^2=-e^{2\phi(r)}dt^2+e^{2\psi(r)}dr^2+r^2d\Omega_{3}^2,
\end{eqnarray}
where $\Omega_{3}$ is a standard $3$ sphere.

In the present study, we pay attention to that the Bumblebee field has a radial vacuum energy expectation because the spacetime curvature has a strong radial variation, on the contrary, the temporal changes are very slow. So the Bumblebee field is supposed to be spacelike($b_\mu b^\mu=$ positive constant) as that
\begin{eqnarray}\label{bu}
b_\mu=\big(0,be^{\psi(r)},0,0,\cdots,0\big),
\end{eqnarray}
where $b$ is a positive constant.
Then the Bumblebee field strength is
\begin{eqnarray}
b_{\mu\nu}=\partial_{\mu}b_{\nu}-\partial_{\nu}b_{\mu},
\end{eqnarray}
whose components are all zero. And their divergences are all zero, i.e.,
\begin{eqnarray}
\nabla^{\mu}b_{\mu\nu}=0.
\end{eqnarray}
From the equation of motion (\ref{motion}), we have
\begin{eqnarray}
b^{\mu}R_{\mu\nu}=0\label{motion2}.
\end{eqnarray}
The gravitational field equations (\ref{bar}) become
\begin{eqnarray}\label{}
G_{\mu\nu}=2\alpha T^{GB}_{\mu\nu}+\mathcal{B}_{\mu\nu}.
\end{eqnarray}

 For the metric (\ref{metric}), the nonzero components of Einstein tensor $G_{\mu\nu}$, the Gauss-Bonnet momentum tensor $T^{GB}_{\mu\nu}$ and the Bumblebee tensor $\mathcal{B}_{\mu\nu}$ are shown below \cite{wiltshire}:

 \begin{eqnarray}
&&G_{00}=\frac{3e^{2\phi-2\psi}}{2r^2}\Big[2(e^{2\psi}-1)+2r\psi'\Big],\\
&&G_{11}=\frac{3}{2r^2}\Big[2(1-e^{2\psi})+2r\phi'\Big],\\
&&G_{22}=e^{-2\psi}\Big[(1-e^{2\psi})+2r(\phi'-\psi')
+r^2(\phi''+\phi'^2-\phi'\psi')\Big],\\
&&G_{ii}=G_{22}\prod^{i-2}_{j=1}\sin^2\theta_j,\\
&&R_{11}=\frac{3}{r}\psi'-(\phi''+\phi'^2-\phi'\psi'),
\end{eqnarray}

$\mathcal{B}_{\mu\nu}$  are
\begin{eqnarray}
&&\mathcal{B}_{00}=\frac{\varrho b^2e^{2\phi-2\psi}}{2r^2}\Big[6-3r\psi'-r^2(\phi''+\phi'^2
-\phi'\psi')\Big],\\
&&\mathcal{B}_{11}=-\frac{\varrho b^2}{2r^2}\Big[6-3r(\psi'-2\phi')+r^2(\phi''+\phi'^2-\phi'\psi')\Big],\\
&&\mathcal{B}_{22}=-\frac{\varrho b^2e^{-2\psi}}{2}\Big[2-r\psi'+4r\phi'+r^2(\phi''
+\phi'^2-\phi'\psi')\Big],\\
&&\mathcal{B}_{ii}=\mathcal{B}_{22}\prod^{i-2}_{j=1}\sin^2\theta_j.
\end{eqnarray}
The nonzero components of the Gauss-Bonnet term $T^{GB}_{\mu\nu}$ are
\begin{eqnarray}
&&T^{GB}_{00}=- \frac{6}{r^2}e^{2\phi-2\psi}
\Big[\frac{2}{r}(1-e^{-2\psi})\psi'\Big],\\
&&T^{GB}_{11}=\frac{6}{r^2}
\Big[-\frac{2}{r}(1-e^{-2\psi})\phi'\Big],\\
&&T^{GB}_{22}=2e^{-2\psi}\Big\{-4e^{-2\psi}\phi'\psi'-2(1-e^{-2\psi})(\phi''+\phi'^2-\phi'\psi')\Big\},\\
&&T^{GB}_{ii}=T^{GB}_{22}\prod^{i-2}_{j=1}\sin^2\theta_j.
\end{eqnarray}

 By using the motion equation (\ref{motion2})
\begin{eqnarray}R_{11}=\frac{3}{r}\psi'-(\phi''+\phi'^2-\phi'\psi')=0,\end{eqnarray}
one can obtain the following three gravitational field equations
\begin{eqnarray}
&&2(e^{2\psi}-1)+2r\psi'=-8\alpha\Big[\frac{2}{r}(1-e^{-2\psi})\psi'\Big]+\lambda\big[2-2r\psi'\big],\label{tt}\\
&&2(1-e^{2\psi})+2r\phi'=8\alpha\Big[-\frac{2}{r}(1-e^{-2\psi})\phi'
\Big]-\lambda\big[2+2r\phi'\big],\label{rr}\\
&&(1-e^{2\psi})+r\psi'+2r\phi'=4\alpha
\Big\{-4e^{-2\psi}\phi'\psi'+(1-e^{-2\psi})\Big[
-\frac{6}{r}\psi'\Big]\Big\}
-\lambda\Big[1+r\psi'+2r\phi')\Big]\label{theta},
\end{eqnarray}
where we denote the Lorentz-violating parameter $\lambda=\varrho b^2$.
By adding Eq. \eqref{tt} to Eq. \eqref{rr}, we derive the result $\phi'(r)+\psi'(r)=0$, as the alternative solution $1+2\lambda+4\alpha(1-e^{-2\psi})/r^2=0$ is inconsistent with the remaining field equations.
If we rewrite the function $e^{2\psi}=(1+\lambda)/f(r)$, then Eq. \eqref{tt} can be transformed into the form:
\begin{eqnarray}\label{diff01}
f'(r) \left((\lambda +1) \left(8 \alpha +(\lambda +1) r^2\right)-8 \alpha  f(r)\right)+2 (\lambda +1)^2 r (f(r)-1)=0.
\end{eqnarray}

The other metric function is connected with the function $f(r)$ as
\begin{eqnarray}
e^{2\phi(r)}=f(r).
\end{eqnarray}

Solving differential equation \eqref{diff01}, we obtain an explicit form of the function $f(r)$ as given by,

\begin{equation} \label{metricfun}
    f(r) = 1+\lambda+\frac{(\lambda +1)^2 r^2}{8 \alpha } \left(1\pm\sqrt{1+\frac{16 \alpha  \lambda }{(\lambda +1)^2 r^2}+\frac{32 \alpha  M}{(\lambda +1)^2 r^4}}\right).
\end{equation}
This is the black hole metric function, obtained in $5D$.
One may note that for the metric function \eqref{metricfun} with $\alpha>0$, there exist two branches of solutions. The first branch corresponds to the $``+"$ sign, while the second corresponds to the $``-"$ sign. For a similar class of black hole solution, Boulware and Deser \cite{boulware} have demonstrated that the positive branch is unstable and results in a graviton ghost. In this case also, the positive branch is unstable and hence we shall focus on the negative branch for our investigation.

\section{Scalar Perturbation and Quasinormal modes} \label{sec04}

In this analysis, we will investigate the higher-order WKB approximation method to determine the quasinormal modes associated with the massless scalar perturbation. The behavior of uncharged and massless scalar fields is governed by the Klein-Gordon equation:
\begin{equation}
\Box \Psi=0,
\end{equation}

where $\Box$ denotes the D'Alembert operator. This equation can be further rewritten as:

\begin{equation}
\frac{1}{\sqrt{-g}}\partial_\mu(\sqrt{-g}g^{\mu \nu}\partial_\nu)\Psi=0,
\end{equation}

where $\sqrt{-g}$ in the above equation, can be expressed for our 5-dimensional metric by:

\begin{equation}
\sqrt{-g}=r^{3} \sqrt{(1+\lambda)}  \prod_{i=1}^{3} \sin \theta_i.
\end{equation}

To proceed further, we intend to separate the radial and angular parts in the Klein-Gordon equation. For this purpose, we utilise the ansatz:

\begin{equation}
\Psi=e^{-i \omega t} \phi(r) Y_{lm}(\Omega),
\end{equation}
where $\omega$ represents the frequency, $l$ is the azimuthal quantum number, and $m$ ranges from $-l$ to $l$ as the spherical harmonic index. By using this assumption, we obtain the radial equation associated with the scalar field as given below,

\begin{equation}
\phi''+\left(\frac{3}{r}+ \frac{f'(r)}{f(r)} \right) \phi'+\left( \frac{\omega^2 (1+\lambda)}{f^2(r)}-\frac{(1+\lambda)(l(l+2))}{f(r) r^2} \right) \phi=0,
\end{equation}
where the prime indicates a derivative with respect to the radial coordinate, $r$. Through the transformation $\phi=\frac{u}{r^{\frac{3}{2}}}$, we can express this equation as:

\begin{equation}
\frac{f^2(r)}{1+\lambda}u''+\frac{f(r) f'(r)}{1+\lambda}u'+\left[\omega^2- \frac{3}{2} \frac{f(r) f'(r)}{1+\lambda} \frac{1}{r}-\frac{3}{4r^2}\frac{f^2(r)}{1+\lambda}-\frac{l(l+2)f(r)}{r^2} \right]u=0.
\end{equation}
To obtain a Schr\"{o}dinger-like wave equation, we further introduce the tortoise coordinate $dr_{*}=\sqrt{1+\lambda} \frac{dr}{f(r)}$, as shown below:

\begin{equation}
\frac{d^2u}{dr_*^2}+\left[\omega^2-V_{s} \right]u=0,
\end{equation}

where the effective potential associated with the scalar potential has the following explicit form:

\begin{equation}
V_{s}=f(r)\left[\frac{3}{4r^2}\frac{f(r)}{1+\lambda}+\frac{3f'(r)}{2(1+\lambda)r}+\frac{l(l+2)}{r^2} \right].
\end{equation}

\begin{figure}
\centering{
\includegraphics[width=4.8 cm]{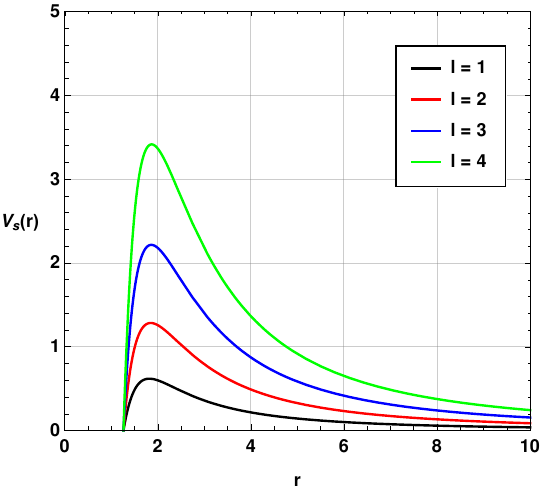}
    \hspace{0.3cm}
   	\includegraphics[width=4.9 cm]{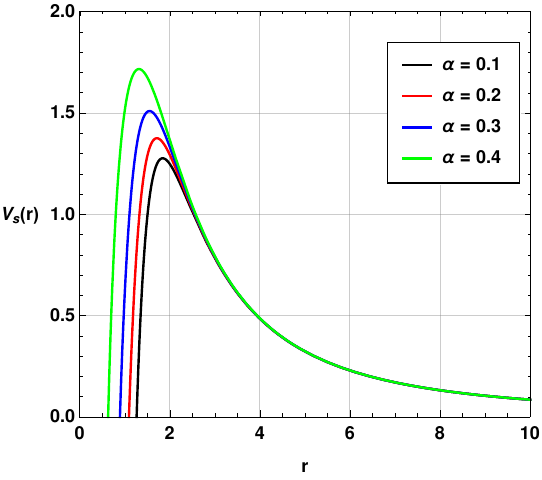}
    \hspace{0.3cm}
\includegraphics[width=4.8 cm]{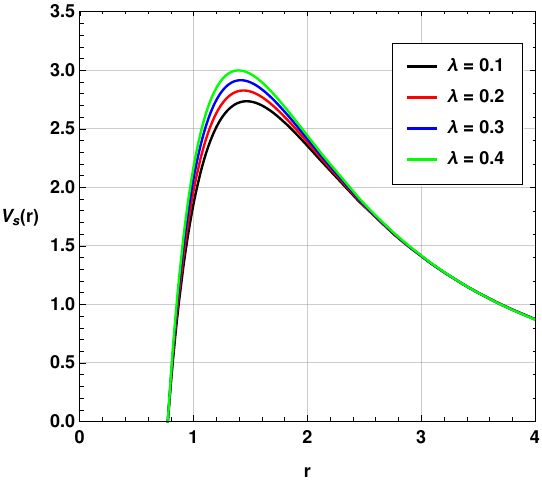}
    }
    \caption{Behaviour of scalar potential in units of black hole mass $M$. On the first panel, we have used $\lambda = 0.1 = \alpha$, on the second panel $\lambda=0.2$ and $l = 2$ and finally on third panel $\alpha = 0.35$ and $l = 3$.} \label{Pot}
\end{figure}

We have shown the variation of the potential in Fig. \ref{Pot}. We observe that the model parameters $\alpha$ and $\lambda$ have significant impacts on the behaviour of the potential. With an increase in the parameter $\alpha$, the peak of the potential shifts towards the event horizon of the black hole and the peak value of the potential also increases significantly. However, the other parameter $\lambda$ impacts the peak of the potential slightly. Since, the peak of the potential is closely associated with the behaviour of the quasinormal modes, we can conclude that the both the model parameters might have noticeable impacts on the quasinormal mode spectrum of the black hole.

\subsection{Pad\'e averaged WKB approximation method} \label{sec04A}

Here, we utilized the sixth-order Pad'e averaged Wentzel-Kramers-Brillouin (WKB) approximation technique to determine the oscillation frequency $\omega$ of GWs using the following expression \cite{Konoplya:2019hlu, Konoplya:2011qq}:
\begin{equation}\label{qnmseqn}
\omega = \sqrt{-\, i \left[ (n + 1/2) + \sum_{k=2}^6 \bar{\Lambda}_k \right] \sqrt{-2 V_0''} + V_0},
\end{equation}

This method has proven to be a reliable tool for calculating the oscillation frequencies of GWs in different astrophysical contexts \cite{Konoplya:2019hlu, Konoplya:2011qq, Konoplya_wkb,Maty_wkb,Cuyubamba:2016cug,Konoplya:2022kld,Zinhailo:2019rwd}. By incorporating the sixth-order WKB method, we were able to improve the accuracy of the calculations, making them more consistent with observational data.

The variable $n$ in Eq. \eqref{qnmseqn} represents overtone numbers and takes on integer values, including $0$, $1$, $2$, and so on. The value of $V_0$ is obtained by evaluating the potential function $V_s$ at the position $r_{max}$, where the potential reaches its maximum value. At this point, the derivative of $V_s$ with respect to $r$ becomes zero. The second derivative of the potential function $V_s$ with respect to $r$, evaluated at the same position $r_{max}$, is represented as $V_0''$.

We also incorporated supplementary correction terms, identified as $\bar{\Lambda}_k$, to improve the accuracy of the calculations. These correction terms are explicitly defined in Ref.s \cite{Schutz,Will_wkb,Konoplya_wkb,Maty_wkb}. They account for higher-order effects, such as mode mixing, and are crucial in accurately predicting the oscillation frequencies of GWs in different astrophysical contexts.

It is important to note that the Pad\'e averaging procedure, coupled with the use of correction terms, greatly improves the accuracy of the calculations. The sixth-order WKB method is an excellent tool for calculating the oscillation frequencies of GWs and can be used to study a wide range of astrophysical phenomena, including black hole mergers, neutron star oscillations, and cosmic string vibrations. Overall, the combination of the sixth-order WKB method and correction terms provides a powerful and accurate technique for predicting the oscillation frequencies of GWs in different astrophysical contexts.

\begin{table}[ht]
\caption{Quasinormal modes of the black hole with $n= 0$, $M=1$, $\alpha=0.1$ and $\lambda = 0.15$ for the massless scalar perturbation.}
\label{QNMtab01}
\begin{center}
{\small 
\begin{tabular}{|cccc|}
\hline
\;\;\;\;$l$ & \;\;\;\; Pad\'e averaged WKB\;\;\;\;
& \;\;$\vartriangle_{rms}$\;\;\;\; & \;\;$\Delta_6$\;\; \\ \hline
 \;\;$l=1$ & $0.73235\, -0.220539 i$ & $0.00125271$ \;\; & $0.00220978$ \;\; \\
 \;\; $l=2$ & $1.09747\, -0.218764 i$ & $0.000353843$ \;\; & $0.000455604$ \;\; \\
 \;\; $l=3$ & $1.46328\, -0.218223 i$ & $0.000113383$ \;\; & $0.000130362$ \;\; \\
  \;\;$l=4$ & $1.82925\, -0.217972 i$ & $0.0000452552$  \;\;& $0.0000890869$  \;\;\\
  \;\;$l=5$ & $2.19525\, -0.217833 i$ & $0.000021469$  \;\;& $0.0000141825$  \;\;\\
  \;\;$l=6$ & $2.56124\, -0.217747 i$ & $0.0000113702$  \;\;& $0.0000118111$ \;\; \\ \hline
\end{tabular}
}
\end{center}
\end{table}

We have shown the numerically calculated quasinormal modes using 6th order Pad\'e averaged WKB method, in Table \ref{QNMtab01}.It presents the quasinormal modes for varying multipole moments, focusing on cases where the overtone number is $n=0$. The second column of the table shows these modes, calculated using the 6th-order Pad\'e averaged WKB approximation method. The table includes the $\Delta_{rms}$ value, indicating the root mean square error of this approximation, and introduces $\Delta_6$ to measure the difference between adjacent order approximations. $\Delta_6$ is computed as $\Delta_6 = \dfrac{|\omega_7 - \omega_5|}{2}$, where $\omega_7$ and $\omega_5$ are the quasinormal modes obtained using the Pad\'e averaged 7th order and 5th order WKB approximation methods, respectively. The error associated with the quasinormal modes decreases as the multipole moment $l$ increases, a typical behavior of the WKB approximation method. However, this method may not provide accurate results when the overtone number $n$ exceeds the multipole moment $l$ \cite{Lambiase:2023hng, Gogoi:2023kjt, Parbin:2022iwt, Gogoi:2022ove, Gogoi:2022wyv, Gogoi:2023fow, Sekhmani:2023ict}.
\begin{figure}[htbp]
\centerline{
   \includegraphics[scale = 0.5]{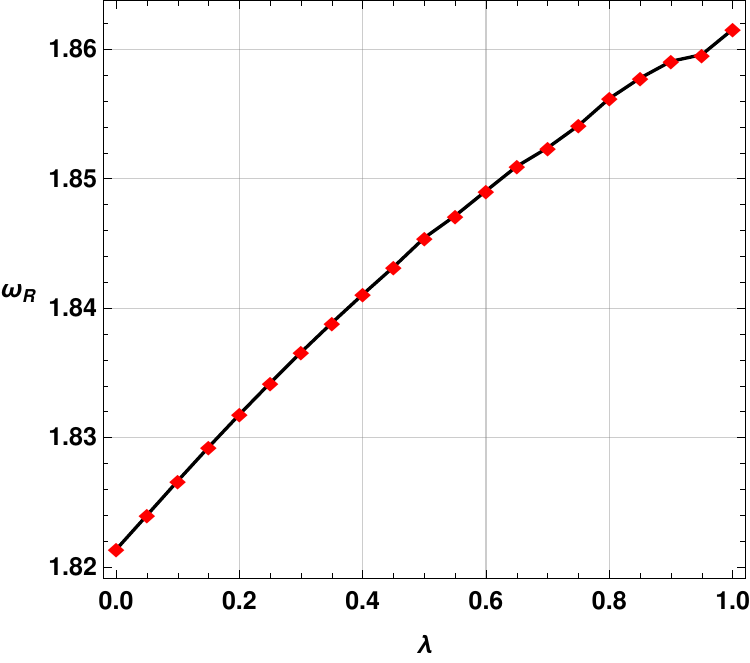}\hspace{0.5cm}
   \includegraphics[scale = 0.51]{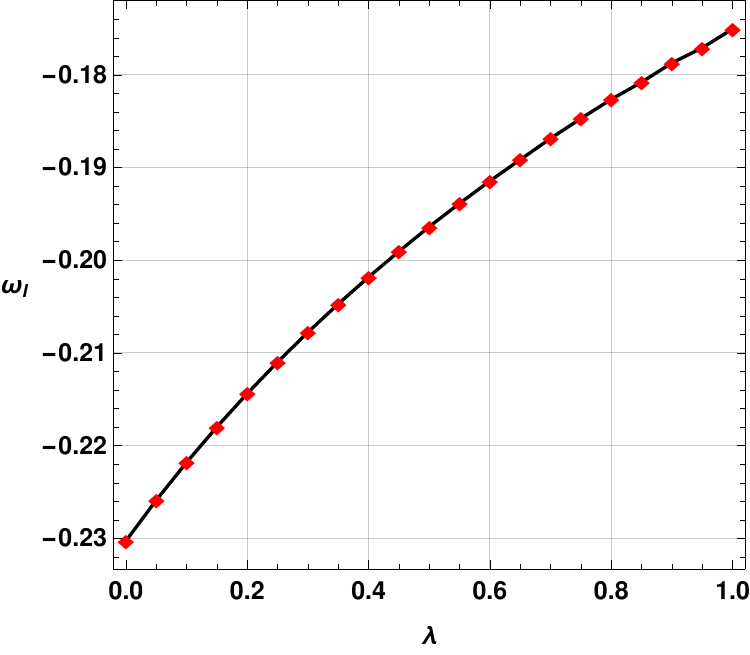}} \vspace{-0.2cm}
\caption{Variation of quasinormal modes with respect to model parameter $\lambda$
with $\alpha=0.1,$ $n= 0$ and $l=4$ for massless scalar perturbation.}
\label{QNMs01}
\end{figure}

\begin{figure}[htbp]
\centerline{
   \includegraphics[scale = 0.5]{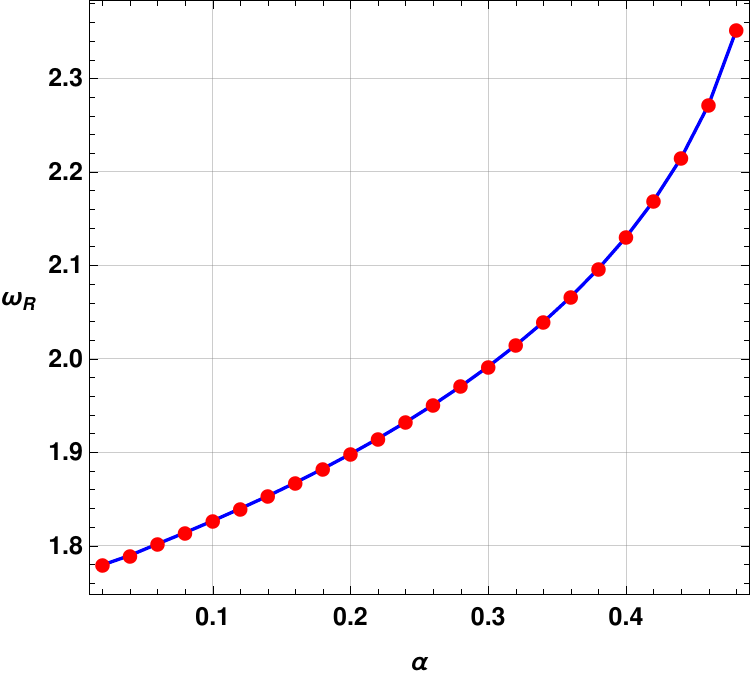}\hspace{0.5cm}
   \includegraphics[scale = 0.51]{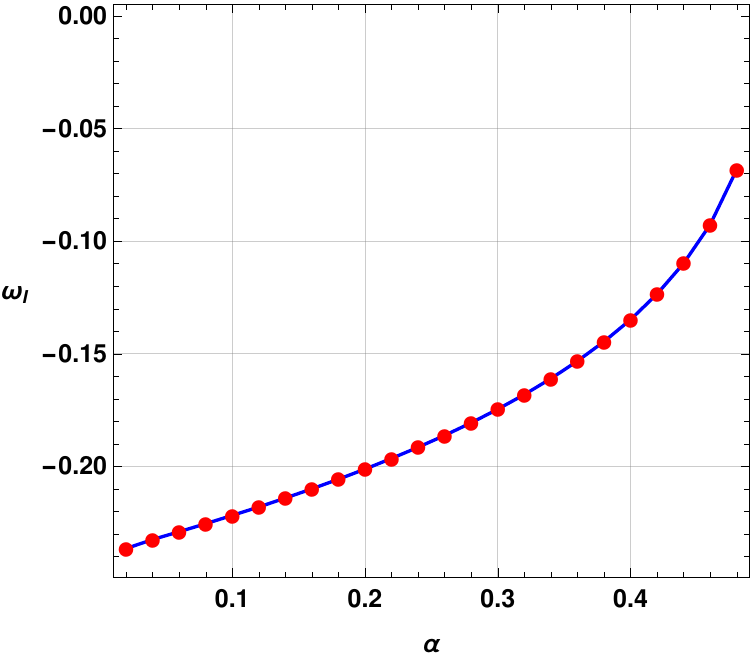}} \vspace{-0.2cm}
\caption{Variation of quasinormal modes with respect to model parameter $\alpha$
with $\lambda=0.1,$ $n= 0$ and $l=4$ for massless scalar perturbation.}
\label{QNMs02}
\end{figure}

We have shown the variations of quasinormal modes with respect to the Lorentz violation parameter $\lambda$ and Gauss-Bonnet coupling parameter $\alpha$ in Fig. \ref{QNMs01} and Fig. \ref{QNMs02} respectively. As seen from the behaviour of the potential, we observe here that the quasinormal mode spectrum is affected more by the coupling parameter $\alpha$ than the Lorentz violation parameter $\lambda$. With an increase in the parameter $\lambda$ from $0$ to $1$, we observe an increase in the real quasinormal mode i.e. ring-down GW frequency while in the case of damping rate, we observe a decrease. The variations are very close to linear. However, in the case of model parameter $\alpha$ also, we observe a similar behaviour. But here the variations are non-linear and more significant.

Our investigation implies that both the Lorentz violation parameter and the Gauss-Bonnet coupling parameter have different impacts on the ring-down GWs and in the near future, with GW observational data from a more sensitive detector like LISA, it might be possible to constrain both the parameters from quasinormal modes.

\subsection{Validity of Hod's conjecture}\label{sec04B}
In the field of black hole physics, one of the most intriguing questions is how the properties of a black hole, such as its mass, charge, and angular momentum, manifest in the black hole's quasinormal modes. In a seminal paper \cite{Hod:2006jw}, Hod proposed that in the spectrum of any black hole's quasinormal modes there always exists a frequency $\omega$ such that the imaginary part of $\omega$ is bounded by $\pi T_H$, where $T_H$ is the Hawking temperature of the black hole, as expressed in the following equation:

\begin{equation}\label{Hod}
\left|Im\left(\omega\right) \right|\leq \pi T_H\,,
\end{equation}
where $T_H$ is the Hawking temperature. The Hawking temperature of a black hole is related to the surface gravity at the event horizon, $r_+$, which can be expressed in terms of the metric function $f(r)$ as $\kappa_H=\frac{1}{2}\frac{d}{dr}\left(f(r)\right)\bigg|_{r=r_+}$. Using the relation between surface gravity and temperature, following equation expresses the Hawking temperature of the black hole in terms of the metric function evaluated at the event horizon as given by
	\begin{equation}\label{TH}
	T_H = \frac{f'(r)}{4 \pi}\Bigg\vert_{r=r_+} = \frac{(\lambda +1) \sqrt{M-2 \alpha }}{2 \sqrt{2} \pi  (-2 \alpha  \lambda +2 \alpha +\lambda  M+M)}.   
	\end{equation}

Remarkably, Hod's claim is general and applies to not only four-dimensional black holes but also to higher-dimensional black holes and even asymptotically Anti-de Sitter (AdS) black holes \cite{Malybayev:2021lfq, Berti:2003jh, Ghosh:2005aq, Churilova:2019sah}.

In summary, Hod's statement (\ref{Hod}) provides a powerful constraint on the quasinormal modes of black holes, and it applies to a wide range of black hole geometries. The Hawking temperature, which is related to the metric function at the event horizon, plays a crucial role in this constraint and determines the upper bound on the damping rate of the fundamental oscillation of the black hole.

\begin{figure}[htbp]
\centerline{
   \includegraphics[scale = 0.5]{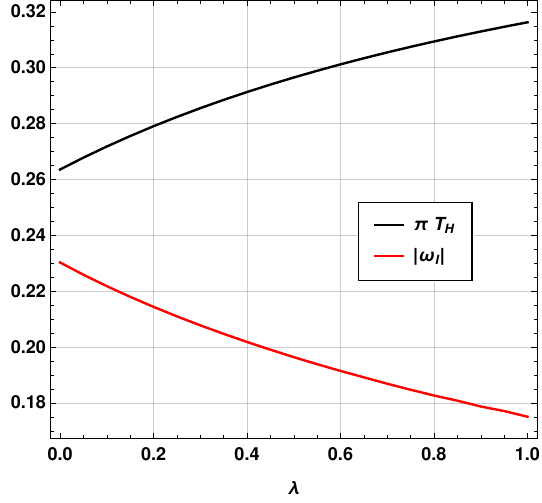}\hspace{0.5cm}
   \includegraphics[scale = 0.51]{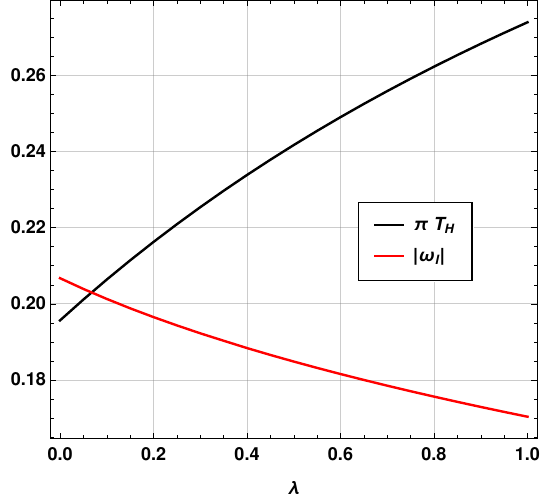}\hspace{0.5cm}
   \includegraphics[scale = 0.51]{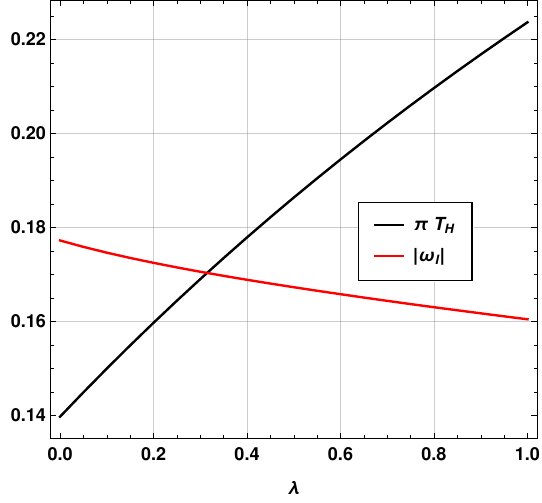}
   } \vspace{-0.2cm}
\caption{Variation of magnitude of imaginary quasinormal modes and $\pi T_H$ with respect to model parameter $\lambda$
with $n= 0$ and $l=4$ for massless scalar perturbation. We have used $\alpha = 0.1, 0.2, 0.3$ on the first, second and third panels, respectively. }
\label{HOD01}
\end{figure}

\begin{figure}[htbp]
\centerline{
   \includegraphics[scale = 0.5]{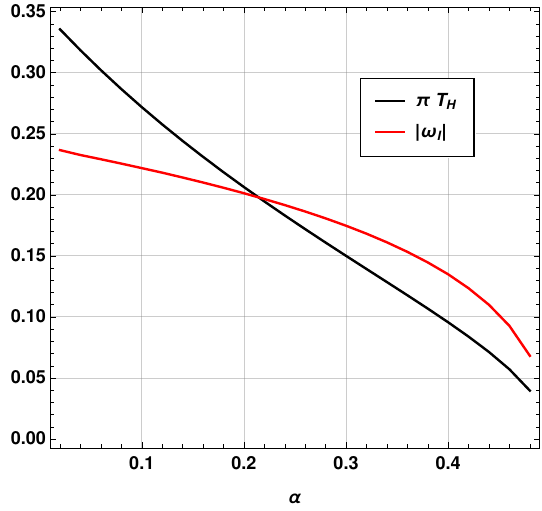}\hspace{0.5cm}
   \includegraphics[scale = 0.51]{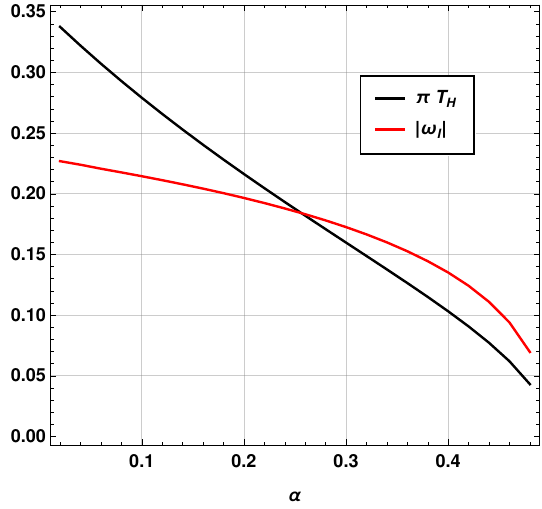}\hspace{0.5cm}
   \includegraphics[scale = 0.51]{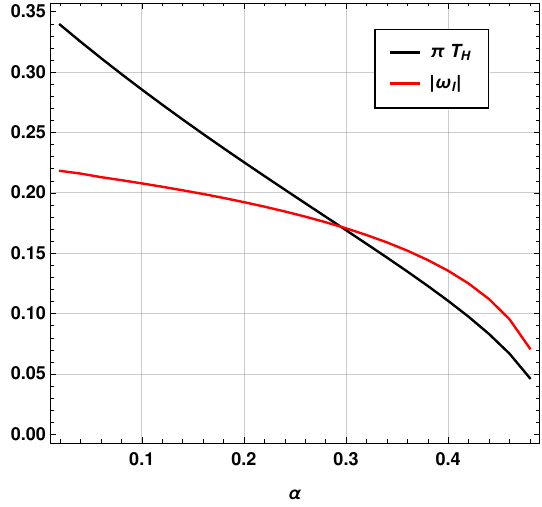}
   } \vspace{-0.2cm}
\caption{Variation of magnitude of imaginary quasinormal modes and $\pi T_H$ with respect to model parameter $\alpha$
with $n= 0$ and $l=4$ for massless scalar perturbation. We have used $\lambda = 0.1, 0.2, 0.3$ on the first, second and third panels, respectively. }
\label{HOD02}
\end{figure}

Hod's conjecture has been extensively studied in the context of black hole physics \cite{Cuyubamba:2016cug,Konoplya:2022kld,Zinhailo:2019rwd}, and its validity has been confirmed for a wide range of black hole geometries. To illustrate the conjecture, we present Fig. \ref{HOD01}, which shows three panels with variations of the damping rate of the fundamental mode, $\omega_I$, with respect to the parameter $\lambda$ for different values of $\alpha$. The black curve represents the upper bound given by Hod's conjecture, i.e., $\pi T_H$, and the red curve represents the damping rate of the fundamental mode, $\omega_I$. In the first panel, the Hod conjecture is satisfied for all ranges of $\lambda$. In the second panel, the conjecture is violated for small $\lambda$ values. Here we observe a threshold point near $\lambda=0.06$ corresponding to $\alpha=0.2$. In the third panel, the intersection point of the black and red curves shifts towards higher $\lambda$ values, indicating that the conjecture may be violated for higher values of $\alpha$. One may note that higher value of $\lambda$ is favourable for the validation of the conjecture i.e. presence of Lorentz violation may indicate the validity of Hod's conjecture. To see the impacts of $\alpha$, we present Fig. \ref{HOD02}, which shows three panels with variations of the damping rate of the fundamental mode, $\omega_I$, with respect to the parameter $\alpha$ for different values of $\lambda$. For smaller values of $\alpha$ and $\lambda$, in all three panels of Fig. \ref{HOD02}, the Hod conjecture is satisfied. In these panels, we see that there is a threshold point of $\alpha$ beyond which the conjecture is violated. This threshold point moves towards higher values of $\alpha$ as $\lambda$ increases. These results suggest that the validity of Hod's conjecture may depend on the specific black hole geometry and its parameters, and higher values of $\alpha$ may lead to violations of the conjecture. In summary, we observed that both the parameters have opposite impacts on the validity of hod's conjecture. Higher values of Lorentz violation parameter $\lambda$ favours the validity of the Hod's conjecture while smaller values of Gauss-Bonnet coupling parameter $\alpha$ favours the validity of Hod's conjecture.

\section{Optical properties of the black hole} \label{sec05}

This section delves into the fascinating and complex phenomena of black hole shadows and their associated emission rates, shedding light on the mysterious nature of these cosmic behemoths. Black holes, characterized by their incredibly powerful gravitational pull, create an environment where even light cannot escape once it crosses the event horizon \cite{EslamPanah:2020hoj, Sekhmani:2024fjn, Gogoi:2024vcx, Gogoi:2023ffh}. This unique property gives rise to what is known as the black hole shadow, a darkened region in space-time that envelops the event horizon. The shadow's appearance is distinct, contrasting against the background of surrounding matter and light, making it a key feature for astronomers to study.

The size and configuration of the black hole shadow, along with its emission rate, provide crucial insights into the fundamental nature of black holes and the laws of gravity. The study of these shadows has been a topic of great interest and importance in astrophysics, as they offer a glimpse into the otherwise hidden properties of these enigmatic objects. Recent advancements in astronomical observation and imaging technology have made it possible to capture detailed images of these elusive black hole shadows, marking a significant milestone in our understanding of these cosmic phenomena \cite{Gogoi:2023ntt, Gogoi:2024vcx, Gogoi:2023ffh}.

The study of black hole shadows has been a focal point for researchers, as these shadows serve as a unique probe into the spacetime geometry around black holes. By analyzing the size and shape of these shadows, one can infer valuable information about the mass, spin, scalar hairs and other properties of the black hole. Furthermore, the emission rate of the black hole, which is closely related to its temperature and hairs, provides insights into the structure of the black hole and the release of energy in the form of radiation.

\subsection{Shadow radius}\label{sec05A}

In this study, we explore the relationship between the black hole shadow and the real part of quasinormal modes frequencies. To establish this connection, we first examine the shadow of the 5D black hole discussed above using the geodesic approach. Here the spacetime metric \eqref{metric} exhibits symmetry, leading to the existence of three Killing vectors: $\partial_t$, $\partial_\phi$, and $\partial_\psi$. These vectors give rise to conserved quantities, namely, the energy $E$, and the angular momenta $L_{\phi}$ and $L_{\psi}$ in the $\phi$ and $\psi$ directions, respectively. Utilizing these conserved quantities and their relations with conjugate momenta, we derive the geodesic equations:

\begin{eqnarray}
	\frac{d t}{d\sigma} &=& \frac{E}{f(r)}, \notag\\ [2mm]
	\frac{d \phi}{d\sigma} &=& \frac{L_{\phi}}{r^2 \sin^2 \theta}, 
	\notag\\ [2mm]
	\frac{d \psi}{d\sigma} &=& \frac{L_{\psi}}{r^2 \cos^2 \theta},
	\label{eom}
\end{eqnarray}
where $\sigma$ represents an affine parameter. Furthermore, the radial and angular geodesic equations can be obtained by employing the Hamilton-Jacobi equation:
\begin{eqnarray}
	\frac{\partial S}{\partial \sigma} = - \frac{1}{2}g^{\mu\nu} 
	\frac{\partial S}{\partial x^\mu} \frac{\partial S}{\partial x^\nu},
	\label{hmj}
\end{eqnarray}
 where $S$ denotes the Jacobi action which can be further 
written as
\begin{eqnarray}
	S = \frac{1}{2} m_0^2 \sigma - E t + L_{\phi} \phi + L_{\psi} \psi 
	+ S_{r}(r) + S_\theta (\theta), \label{jac}
\end{eqnarray}
 The parameter $m_0$ represents the mass of the test particle, which becomes zero in the case of a photon. Substituting equation \eqref{jac} into equation \eqref{hmj} and performing straightforward calculations, we derive the following expressions:
\begin{eqnarray}
	r^2 \frac{d r}{d\sigma} &=& \pm \sqrt{\mathcal{R}(r)}, 
	\notag \\ [2mm]
	r^2 \frac{d \theta}{d\sigma} &=& \pm \sqrt{\Theta(\theta)}, 
	\label{r-th-eom} 
\end{eqnarray}
where $\mathcal{R}(r)$ and $\Theta(\theta)$ are given by:
\begin{eqnarray}
	\mathcal{R}(r) &=& (1+\lambda)^{-1} E^2 r^4 -r^2 f(r) (1+\lambda)^{-1} 
	\left(\mathcal{K} + L_{\phi}^2 + L_{\psi}^2\right), 
	\notag \\ [2mm]
	\Theta(\theta) &=& \mathcal{K} - L_{\phi}^2 \cot^2 \theta 
	- L_{\psi}^2 \tan^2 \theta, \label{R-Th}
\end{eqnarray}
where $\mathcal{K}$ is the Carter constant, which arises during the separation of the coefficients of $r$ and $\theta$ in the Hamilton-Jacobi formulation.

To describe the black hole shadow, we introduce celestial coordinates, which are defined for 5D black holes as \cite{Amir:2017slq}:

\begin{eqnarray}
	x &=& -\lim_{r_0 \rightarrow \infty} r_0
	\frac{p^{\hat{\phi}}+p^{\hat{\psi}}}{p^{\hat{t}}}, \notag \\ [2mm]
	y &=& \lim_{r_0 \rightarrow \infty}	r_0
	\frac{p^{\hat{\theta}}}{p^{\hat{t}}}, \label{celest}
\end{eqnarray}
where $p^{\hat{i}}$ represents the contravariant components of the momenta in the new coordinate basis. These contravariant components can be computed using the orthonormal basis vectors for the local observer \cite{Johannsen:2013vgc,Amir:2017slq}, yielding:
\begin{eqnarray}
	p^{\hat{t}} &=& \frac{E}{f(r)}, \quad 
	p^{\hat{\phi}} = \frac{L_{\phi}}{r \sin \theta}, \quad
	p^{\hat{\psi}} = \frac{L_{\psi}}{r \cos \theta}, \notag \\ [2mm]
	p^{\hat{r}} &=& \pm \sqrt{f(r) \mathcal{R}(r)},\quad
	p^{\hat{\theta}} = \pm \frac{\sqrt{\Theta(\theta)}}{r}. \label{thrmom}
\end{eqnarray}

Inserting \eqref{thrmom} into \eqref{celest} and taking the limit as $r_0 \rightarrow \infty$, we obtain:
\begin{eqnarray}
	x &=& -\left( \xi_{\phi} \csc \theta +\xi_{\psi} \sec \theta \right), 
	\notag \\ [2mm]
	y &=& \pm \sqrt{\eta -\xi_{\phi}^2 \cot^2 \theta 
	-\xi_{\psi}^2 \tan^2 \theta}, 
	\label{celest}
\end{eqnarray}
where we introduce the new quantities, $\xi_{\phi} =L_{\phi}/E$, $\xi_{\psi} =L_{\psi}/E$, and $\eta =\mathcal{K}/E^2$. These quantities are referred to as the impact parameters. Now, let's consider the limiting cases of \eqref{celest} based on the observer's location. If the observer is in the equatorial plane ($\theta_0 = \pi/2$), then $L_{\psi} = 0$, implying $\xi_{\psi} = 0$. This results in:
\begin{eqnarray}
	x = -\xi_{\phi} , \quad	y = \pm \sqrt{\eta }.
	\label{celestcord1}
\end{eqnarray}
Conversely, when $\theta_0 = 0$, we have $L_{\phi} = 0$, leading to $\xi_{\phi} = 0$:
\begin{eqnarray}
	x = -\xi_{\psi} , \quad	y = \pm \sqrt{\eta}.
	\label{celestcord2}
\end{eqnarray}
Equations \eqref{celestcord1} and \eqref{celestcord2} establish the connection between the celestial coordinates and the impact parameters for observers positioned at inclination angles $\theta_0=\pi/2$ and $\theta_0=0$, respectively \cite{Das:2019sty}. These equations are fundamental for extracting information regarding the shadows of black holes \cite{Sekhmani:2024fjn, Gogoi:2024vcx, Gogoi:2023ffh, Vagnozzi:2022moj}.

\begin{figure}
\centering{
   	\includegraphics[width=7.8 cm]{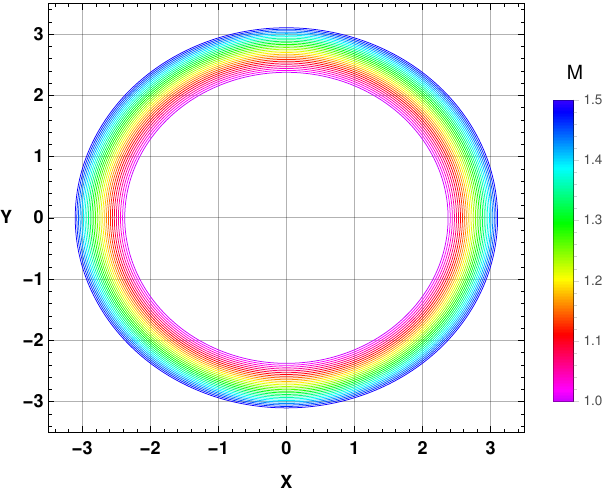}
    }
    \caption{Stereographic projection of black hole shadow with $\alpha = 0.3$ and $\lambda = 0.5.$ \label{rs01}}
\end{figure}

\begin{figure}
\centering{
   	\includegraphics[width=7.8 cm]{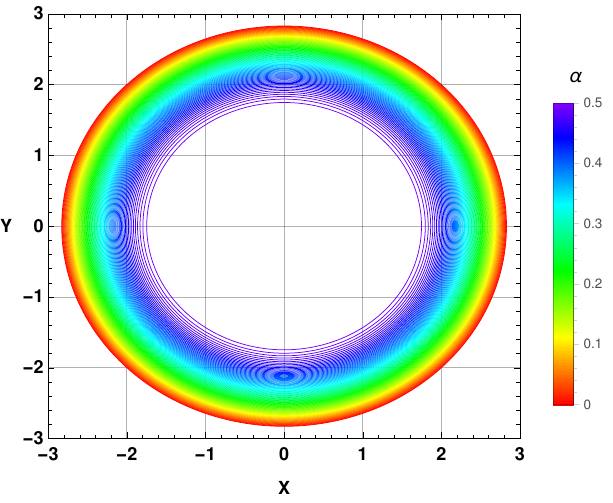}
    \hspace{0.5cm}
\includegraphics[width=7.8 cm]{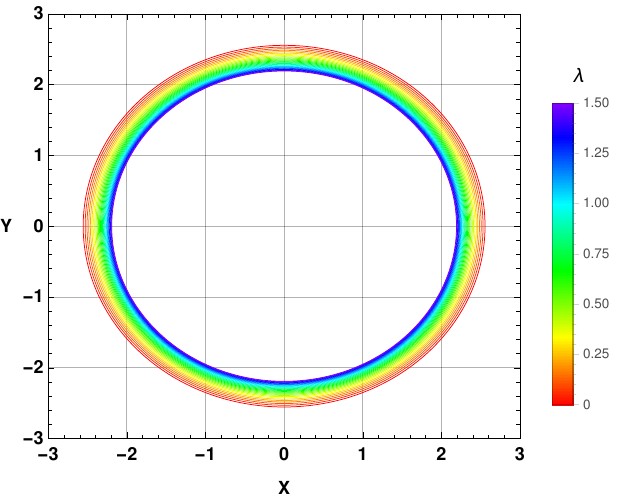}
    }
    \caption{Stereographic projection of black hole shadow. On the first panel, we have used $M = 1$ and $\lambda = 0.5$, and on the second panel, we have used $M = 1$ and $\alpha = 0.3$. \label{rs02}}
\end{figure}

We further establish a connection between the impact parameters of the metric \eqref{metric} by utilizing the conditions for unstable spherical photon orbits, $\mathcal{R}=0$ and $d\mathcal{R}/dr=0$, resulting in:

\begin{equation}\label{photonsphere}
    r = \frac{2 f(r)}{f^{\prime}(r)}\Bigg\vert_{r=r_{ph}},
\end{equation}
and
\begin{equation}\label{shadoweq}
\eta + \xi_{\phi}^2 + \xi_{\psi}^2 = \frac{r^2}{f(r)}\Bigg\vert_{r=r_{ph}},
\end{equation} 
where the prime ($\prime$) indicates differentiation with respect to $r$. In the equations above, $r_{ph}$ denotes the photon sphere, which can be determined by solving \eqref{photonsphere}. Equation \eqref{shadoweq} reveals that $\frac{r_{ph}^2}{f(r_{ph})}=R_s^2$, where $R_s$ represents the shadow radius of the black hole. A stereographic projection of the black hole shadow can be obtained from Eqs. \eqref{celest} based on this shadow radius. We have shown the stereographic projections of the black hole shadows in Figs. \ref{rs01} and \ref{rs02}.


In Fig. \ref{rs01}, we show the variation of the shadow radius for different black hole masses. In Fig. \ref{rs02}, we show the impacts of the model parameters $\alpha$ and $\lambda$ on the black hole shadow. In both cases, with an increase in the model parameters, the shadow gets decreased. The impact of $\alpha$ on black hole shadows is more significant than the impact of $\lambda$.

\subsection{Emission rate}\label{sec05B}
The particle emission rate from a black hole is another optical property of the black hole. The emission rate of the black hole can give a good measure of its stability and lifetime. It is mainly associated with two significant properties of the black hole directly. They are the Hawking temperature and the shadow. The expression of energy emission rate reads \cite{Wei2013,emission2, Cai:2020igv, Decanini:2011xi}
	
\begin{equation} \label{emissioneq}
\frac{d^2 E(\bar{\omega})}{d\bar{\omega} dt}=\frac{2\pi^2 \sigma_{lim}}{exp\Big({\frac{\bar{\omega}}{T_{H}}}\Big)-1}\bar{\omega}^3
\end{equation}
where $ E(\bar{\omega}) $ and $\bar{\omega}$ stand for the energy and frequency of emission corresponding to the black hole. 
	
The  expression  for $\sigma_{lim}$, which is  the  limiting  constant  value is expressed in $D$ spacetime dimensions as \cite{Cai:2020igv, Decanini:2011xi}
\begin{equation}
\sigma_{lim}=\frac{\pi^{\frac{D-2}{2}}R_{s}^{D-2}}{\Gamma \left( \frac{D}{2} \right)}
\end{equation}
where $R_s$ is the radius of the shadow. In $D=5$ dimensions, $\sigma_{lim}$ reads
\begin{equation}
\sigma_{lim}\approx\frac{4\pi R_s^3}{3}~.
\end{equation}

Finally, using $\sigma_{lim}$ in \eqref{emissioneq}, the energy emission rate is expressed as follows \cite{Wei2013,Cai:2020igv, Decanini:2011xi}:

\begin{equation} \label{emission_final}
    \frac{d^2 E(\bar{\omega})}{dt d \bar{\omega} }= \frac{8 \pi^3 \bar{\omega}^3 R_s^3 }{3 \left\lbrace\exp(\bar{\omega}/T_H ) - 1\right\rbrace}.
\end{equation}

\begin{figure}
\centering{
   	\includegraphics[width=7.8 cm]{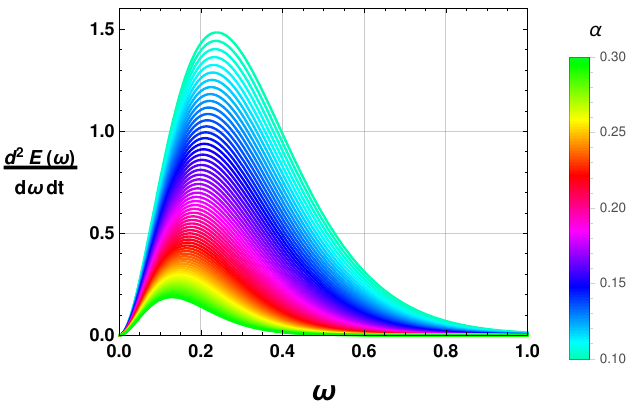}
    \hspace{0.5cm}
\includegraphics[width=7.8 cm]{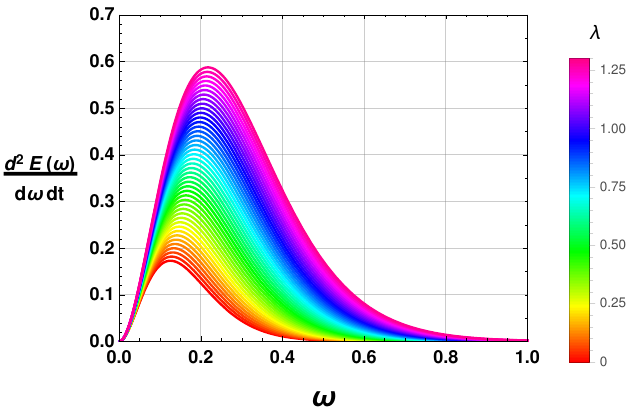}
    }
    \caption{Emission Behaviour of the black hole with $M = 1$. On the first panel, we have used $\lambda = 0.03$ and on the second panel, we have used $\alpha = 0.3.$ \label{fig_emission}}
\end{figure}

We have plotted the emission rate of the black hole using \eqref{emission_final} in Fig. \ref{fig_emission}. On the first panel of Fig. \ref{fig_emission}, we have shown the variation of the emission rate for different values of the parameter $\alpha$. One can see that with an increase in the value of $\alpha$, the emission rate of the black hole decreases resulting a decreased evaporation rate. It implies that with an increase in the parameter $\alpha$, the black hole becomes more stable and its lifetime increases. On the second panel of Fig. \ref{fig_emission}, we have depicted the emission rate for different values of the Lorentz violation parameter $\lambda$. One can see that in presence of Lorentz violation, the emission rate of the black hole increases. It implies that Lorentz violation increases the evaporation rate of the black hole decreasing its lifetime.

From the perspectives of black hole emission rate, we have seen that both the model parameters {\it viz.,} the Gauss-Bonnet term, $\alpha$ and the Lorentz violation term, $\lambda$ have opposite impacts on the evaporation rate of the black hole. The presence of the Gauss-Bonnet term increases the stability of the black hole while the presence of the Lorentz violation term decreases the stability of the black hole.

\subsection{Probing the Hod's conjecture violation via optical properties of the black hole}\label{sec05B}
 Here we shall relate the imaginary quasinormal modes of the black hole with the optical properties discussed above. One may note that the connection of black hole shadow with the real quasinormal modes from a black hole in the eikonal limit has been studied earlier \cite{Cuadros-Melgar:2020kqn}. Here, rather than the real quasinormal modes, we shall focus on the damping rate of quasinormal modes.

For the mathematical feasibility, here we shall consider up to $3$rd order expansion of quasinormal mode expression only, which reads \cite{Cuadros-Melgar:2020kqn}

\begin{eqnarray}
\nonumber
\omega = \Bigg\{  V_s + \frac{V_4}{8V_2}\left( \nu^2+\frac{1}{4}\right) - \left( \frac{7 + 60\nu^2}{288}\right) \frac{V_3^2}{V_2^2}+i\nu \sqrt{-2V_2}\left[\frac{1}{2V_2}\left[\frac{5V_3^4(77+188\nu^2)}{6912V_2^4} \right. \right. \\
\nonumber
\left. \left. -\frac{V_3^2V_4(51+100\nu^2 )}{384V_2^3}+ \frac{V_4^2(67+68\nu^2 )}{2304V_2^2}+\frac{V_5V_3(19+28\nu^2 )}{288V_2^2}+\frac{V_6(5+4\nu^2 )}{288V_2}\right] -1 \right]  \Bigg\}^{1/2}_{r=r_{ph}}.\\
\label{eq71}
\end{eqnarray}
In this expression, $V_i$ represents the {\it i-th} derivative of the potential $V_s$ and $\nu = n+\frac{1}{2}$. Considering $3$rd order formula does not affect the results at the eikonal limit as it produces the same expansion for the first terms of the eikonal
limit when compared to the $4^{th}$ to $6^{th}$ order representation. The quasinormal frequency $\omega$ is calculated at $r_0$, which is the maximum value of the potential $V_s$ calculated by using \cite{Gogoi:2024vcx} \\

\begin{equation}
\label{e8}
\frac{dV_s}{dr}\Bigg|_{r=r_0}= f(r)\left[\frac{3}{4r^2}\frac{f(r)}{1+\lambda}+\frac{3f'(r)}{2(1+\lambda)r}+\frac{l(l+2)}{r^2} \right] \Bigg|_{r=r_0}= 0.
\end{equation}
The latter equation renders different values of $r_0$ depending on
the physical field, theory, and black hole (expressed through $g$),
but to leading - and first sub-leading - order its solution at the eikonal limit is the very simple relation, 
\begin{equation}
\label{e9}
\frac{d}{dr}\left(\frac{f(r)}{r^2}\right)  \Bigg|_{r=r_0}= 0.
\end{equation}
The interesting fact is that $r_0$ and $r_{ps}$ represent the same point, defining $G = \frac{f(r)}{r^2}$, those equations can be written in the form
\begin{equation}
\label{ex1}
\frac{d}{dr}\left(G\right)  \Bigg|_{r=r_0}= 0, \hspace{2.0cm} \frac{d}{dr}\left(G^{-1}\right)  \Bigg|_{r=r_{ps}}= 0.
\end{equation}
As a consequence, $r_{ps}=r_0$, as long as $G^{-2}|_{r=r_{ps}} \neq 0$, which is the case in general.

 This result states that for every spherically symmetric black hole that possesses a photon sphere, the position of the maximum of the potential of motion equations of fields corresponds to the stability threshold for the circular null geodesic around the structure. 

Finally, by expanding the relation (\ref{eq71}) and considering the real part only, we obtain at the eikonal regime, 
\begin{equation}
\label{eqreal}
\omega_R = R_s^{-1}\left(\ell + 1 +\mathcal{O}(\ell^{-1})\right).
\end{equation}

This relation connects the shadow and the quasinormal frequency of the black hole.
By comparing the imaginary terms in the eikonal regime,
\begin{equation}
\label{eqimg}
 \omega_I = \frac{n +1}{\sqrt{2}}R_s^{-1}\sqrt{2f-r^2f''}\Bigg|_{r=r_{ps}} +\mathcal{O}(\ell^{-1}),
\end{equation}

We have already seen that although the black hole shadow does not have a direct correlation with the violation of Hod's conjecture, the emission rate shows a correspondence. The above relation shows that the additional term with $R_s$ i.e. $\sqrt{2f-r^2f''}\Bigg|_{r=r_{ps}}$ plays a significant role.

\section{Conclusion}\label{section6}
In this work, quasinormal modes of a 5 dimensional black hole in the framework of Gauss-Bonnet Bumblebee theory of gravity have been investigated by using higher order WKB approximation method. It is found that the Gauss-Bonnet coupling parameter $\alpha$ affects the quasinormal mode spectrum non-linearly. On the other hand, the Lorentz violation parameter $\lambda$ has a smaller impact on the quasinormal mode spectrum in comparison to the Gauss-Bonnet coupling term $\alpha$. The impact of $\lambda$ on the quasinormal mode spectrum is close to linear. In the next part, we investigated the validity of Hod's conjecture in the Gauss-Bonnet Bumblebee theory of gravity framework. Our investigation into Hod's conjecture within the context of black hole physics has provided valuable insights into the behaviour of Gauss-Bonnet Bumblebee black holes under various conditions. Through our analysis, we have observed that Hod's conjecture holds true for a wide range of black hole geometries, as evidenced by the agreement between the predicted upper bound for the damping rate of the fundamental mode. However, our study also revealed instances where Hod's conjecture is violated, particularly for small values of the Lorentz violation parameter $\lambda$ and higher values of Gauss-Bonnet coupling parameter $\alpha$. 

One of the key findings of our study is the correlation between the validity of Hod's conjecture and the values of $\lambda$ and $\alpha$. We observed that higher values of $\lambda$ tend to support the validity of Hod's conjecture, while smaller values of $\alpha$ favour its validation. This suggests that the specific characteristics of black hole geometries and their parameters play a crucial role in determining the applicability of Hod's conjecture.

We also investigated the optical properties of the black hole {\it viz.} shadow and emission rate. With an increase in both of the model parameters $\lambda$ and $\alpha$, we observe a decrease in the shadow of the black hole. However, in the case of emission rate, with an increase in the Gauss-Bonnet coupling parameter $\alpha$, the emission rate decreases. The presence of Lorentz violation, on the other hand, increases the evaporation rate of the black hole and as a result, the lifetime decreases. Although black hole shadow does not show a clear correlation with the violation of Hod's conjecture, in the case of emission rate, we observe a correlation with the Hod's conjecture. It is seen that a black hole with higher evaporation rate and smaller lifetime favours the Hod's conjecture. In the case of a black hole with a higher lifetime and smaller evaporation or emission rate, the Hod's conjecture is likely to be violated.

Overall, our research contributes to the ongoing discourse in black hole physics, highlighting the intricate relationship between Lorentz violation, Gauss-Bonnet coupling, and the validity of Hod's conjecture along with the optical properties of the black hole. Further studies incorporating additional parameters and exploring different black hole geometries are warranted to gain a more comprehensive understanding of these phenomena.


\begin{thebibliography}{99}
\bibitem{kost2004}V. A. Kosteleck\'y, 
{Phys. Rev. D {\bf 69}, 105009 (2004)} [arXiv:hep-th/0312310].

\bibitem{casana2018}R. Casana, A. Cavalcante, F. P. Poulis, and E. B. Santos, 
{Phys. Rev. D {\bf 97}, 104001 (2018)} [arXiv:1711.02273].

\bibitem{bluhm2005}R. Bluhm and V. A. Kosteleck\'y, 
{Phys. Rev. D {\bf 71}, 065008 (2005)} [arXiv:hep-th/0412320].

\bibitem{bailey2006}Q. G. Bailey and V. A. Kosteleck\'y, 
{Phys. Rev. D {\bf 74}, 045001 (2006)} [arXiv:gr-qc/0603030].

\bibitem{bailey2009}Q. G. Bailey, 
{Phys. Rev. D {\bf 80}, 044004 (2009)} [arXiv:0904.0278].

\bibitem{tso2011}R. Tso and Q. G. Bailey, 
{Phys. Rev. D {\bf 84}, 085025 (2011)} [arXiv:1108.2071].

\bibitem{kost2009}V. A. Kosteleck\'y and J. D. Tasson, 
{Phys. Rev. Lett. {\bf 102}, 010402 (2009)} [arXiv:0810.1459].

\bibitem{maluf2013}R. V. Maluf, V. Santos, W. T. Cruz, and C. A. S. Almeida, 
{Phys. Rev. D {\bf 88}, 025005 (2013)} [arXiv:1304.2090].

\bibitem{maluf2014}R. V. Maluf, C. A. S. Almeida, R. Casana, and M. M. Ferreira, 
{Phys. Rev. D {\bf 90}, 025007 (2014)} [arXiv:1402.3554].

\bibitem{santos2015}A. F. Santos, W. D. R. Jesus, J. R. Nascimento, and A. Yu. Petrov, 
{Mod. Phys. Lett. A {\bf 30}, 1550011 (2015)} [arXiv:1407.5985].

\bibitem{kost2016}V. A. Kosteleck\'y, A. C. Melissinos, and M. Mewes, 
{Phys. Lett. B {\bf 761}, 1 (2016)} [arXiv:1608.02592].

\bibitem{kost2016_2}V. A. Kosteleck\'y and M. Mewes, 
{Phys. Lett. B {\bf 757}, 510 (2016)} [arXiv:1602.04782].

\bibitem{kumar2021}S. Kumar Jha, H. Barman, and A. Rahaman, 
{J. Cosmol. Astropart. Phys. {\bf 04}, 036 (2021)} [arXiv:2012.02642].

\bibitem{kanzi2019}S. Kanzi and \.I. Sakall\i, 
{Nuclear Physics B {\bf 946}, 114703 (2019)}.

\bibitem{jha2022}S. K. Jha, S. Aziz, and A. Rahaman, 
{Eur. Phys. J. C {\bf 82}, 106 (2022)} [arXiv:2103.17021].

\bibitem{gullu}\.I. G\"ull\"u and A. \"Ovg\"un, 
{Annals of Physics {\bf 436}, 168721 (2022)}.


\bibitem{Jusufi19}A. \"Ovg\"un, K. Jusufi, and \.{I} Sakall{\i}, 
{Phys. Rev. D {\bf 99}, 024042 (2019)} [arXiv:1804.09911 [gr-qc]].

\bibitem{Pavluchenko:2016wvi}
S.~A.~Pavluchenko,
Phys. Rev. D \textbf{94}, no.2, 024046 (2016)
doi:10.1103/PhysRevD.94.024046
[arXiv:1605.01456 [hep-th]].

\bibitem{wiltshire}D. L. Willtsire, Phys. Lett. B {\bf169}, 36 (1986).
\bibitem{lovelock} D. Lovelock, J. Math. Phys. {\bf12}, 498 (1971); {\it ibid} {\bf13}, 874 (1972).
\bibitem{zwiebach} B. Zwiebach, Phys. Lett. B {\bf156}, 315 (1985).
\bibitem{lin2020} D. Glavan and Chunshan Lin, Phys. Rev. Lett. {\bf 124}, 081301 (2020).
\bibitem{shu} F. Shu, Phys. Lett. B {\bf811}, 135907 (2020); W. Ai,
 Commun. Theor. Phys. {\bf72}, 095402 (2020);  M. Gurses, T. C. Sisman and B. Tekin, Eur. Phys. J. C {\bf80}, 647 (2020);
  S. Mahapatra, Eur. Phys. J. C {\bf80}, 992 (2020);
   M. Gurses, T. C. Sisman and B. Tekin, Phys. Rev. Lett. {\bf125}, 149001 (2020);
  J. Arrechea, A. Delhom and A. Jim\'{e}nez-Cano,  Phys. Rev. Lett. {\bf125}, 149002 (2020).
\bibitem{arrechea}  J. Arrechea, A. Delhom and A. Jim\'{e}nez-Cano,  Chin. Phys. C {\bf45} 013107 (2021).

\bibitem{hennigar}R. A. Hennigar, D. Kubiznak, R. B. Mann and C. Pollack,
 J. High Energy Phys. {\bf07}, 027 (2020); A. Casalino, A. Colleaux, M. Rinaldi and S. Vicentini,
 Phys. Dark Universe, {\bf31}, 100770 (2021); H. Lu and Y. Pang,
 Phys. Lett. B {\bf809}, 135717 (2020); T. Kobayashi, 
J. Cos. Astro. Phys. {\bf07}, 013  (2020).
\bibitem{fernandes}P. G. S. Fernandes, Phys. Lett. B {\bf805}, 135468 (2020).
\bibitem{yang} Ke Yang, Bao-Min Gu, Shao-Wen Wei and Yu-Xiao Liu, Eur. Phys. J. C {\bf80}, 662 (2020).
\bibitem{feng}Jia-Xi Feng, Bao-Min Gu and Fu-Wen Shu, Phys. Rev. D {\bf103}, 064002 (2021).
\bibitem{kumar2004} A. Kumar, S. G. Ghosh, 
    arXiv:2004.01131; A. Kumar, R. Kumar, 
arXiv:2003.13104.

\bibitem{wei2021}S.-W. Wei, Y.-X. Liu, 
 Eur. Phys. J. Plus {\bf136}, 436 (2021); R. Kumar, S. G. Ghosh, 
 JCAP {\bf07}, 053 (2020). 
 
\bibitem{ghosh} S. G. Ghosh, R. Kumar, 
    Class. Quant. Grav. {\bf37},  245008 (2020); 
S. G. Ghosh, S. D. Maharaj, 
 Phys. Dark Univ. {\bf30}, 100687 (2020); 
S. G. Ghosh, S. D. Maharaj, 
Phys. Dark Univ. {\bf31}, 100793  (2021). 


\bibitem{ding2020} C. Ding, C. Liu, R. Casana and A. Cavalcate, Eur. Phys. C {\bf80}, 178 (2020).


\bibitem{li}Z. Li and A. \"{O}vg\"{u}n, Phys. Rev. D {\bf101}, 024040 (2020).

\bibitem{jha} S. K. Jha and A. Rahaman, 
Eur. Phys. J. C {\bf 81}, 345 (2021).

\bibitem{Daghigh2012}R. G. Daghigh and M. D. Green, 
{Phys. Rev. D {\bf 85}, 127501 (2012)} [arXiv:1112.5397 [gr-qc]].




\bibitem{Vishveshwara}C. V. Vishveshwara, 
{Phys. Rev. D {\bf 1}, 2870 (1970)}.

\bibitem{Press}W. H. Press, 
{ApJ {\bf 170}, L105 (1971)}.

\bibitem{Chandrasekhar_qnms}S. Chandrasekhar and S. Detweiler, 
{Proc. R. Soc. Lond. A {\bf 344}, 441 (1975)}.

\bibitem{Ma}C. Ma, Y. Gui, W. Wang, F. Wang, 
{Cent. Eur. J. Phys. {\bf 6}, 194 (2008)} [arXiv:gr-qc/0611146]. 

\bibitem{Pedrotti:2024znu}
D.~Pedrotti and S.~Vagnozzi,
[arXiv:2404.07589 [gr-qc]].

\bibitem{gogoi1}D. J. Gogoi and U. D. Goswami, 
{Eur. Phys. J. C {\bf 80}, 1101 (2020)} [arXiv:2006.04011].

\bibitem{gogoi2}D. J. Gogoi and U. D. Goswami, 
{Indian J. Phys. {\bf 96}, 637 (2022)} [arXiv:1901.11277].

\bibitem{Liang_2017}D. Liang, Y. Gong, S. Hou and Y. Liu, 
{Phys. Rev. D {\bf 95}, 104034 (2017)} [arXiv:1701.05998].

\bibitem{qnm_bumblebee}R. Oliveira, D. M. Dantas, and C. A. S. Almeida, 
{EPL {\bf 135}, 10003 (2021)} [arXiv:2105.07956].

\bibitem{gogoi3}D. J. Gogoi and U. D. Goswami, 
{Physics of the Dark Universe {\bf 33}, 100860 (2021)} [arXiv:2104.13115].

\bibitem{Graca}J. P. M. Gra\c{c}a and I. P. Lobo, 
{Eur. Phys. J. C {\bf 78}, 101 (2018)} [arXiv:1711.08714].

\bibitem{Zhang2}Y. Zhang, Y.X. Gui, F. Li, 
{Gen. Relativ. Gravit. {\bf 39}, 1003 (2007)} [arXiv:gr-qc/0612010]. 

\bibitem{lopez2020}M. Bouhmadi-López, S. Brahma, C.-Y. Chen, P. Chen, and D. Yeom, 
{J. Cosmol. Astropart. Phys. {\bf 07}, 066 (2020)} [arXiv:2004.13061].

\bibitem{Liang2018}J. Liang, 
{Commun. Theor. Phys. {\bf 70}, 695 (2018)}.

\bibitem{Hu}Y. Hu, C.-Y. Shao, Y.-J. Tan, C.-G. Shao, K. Lin, and W.-L. Qian, 
{EPL {\bf 128}, 50006 (2020)}.

\bibitem{hemawati2022}S. Giri, H. Nandan, L. K. Joshi, and S. D. Maharaj, 
{Eur. Phys. J.  Plus {\bf 137}, 181 (2022)}.

\bibitem{gogoi4}D. J. Gogoi, R. Karmakar, and U. D. Goswami, 
{arXiv:2111.00854} (2021).

\bibitem{Gogoi:2022wyv}
D.~J.~Gogoi and U.~D.~Goswami, 
JCAP \textbf{06}, no.06, 029 (2022)
doi:10.1088/1475-7516/2022/06/029
[arXiv:2203.07594 [gr-qc]].

\bibitem{Ding2022}C. Ding, X. Chen, and X. Fu, 
Nuclear Physics B {\bf 975}, 115688 (2022).		



\bibitem{bluhm} R. Bluhm, Shu-Hong Fung and V. A. Kosteleck\'{y}, Phys. Rev. D {\bf77}, 065020
(2008).

\bibitem{Bertolami2005}O. Bertolami and J. P\'aramos, 
{Phys. Rev. D {\bf 72}, 044001 (2005)}.

\bibitem{boulware} D. G. Boulware and S. Deser, Phys. Rev. Lett. {\bf55}, 2656 (1985).

\bibitem{Konoplya:2019hlu}
R.~A.~Konoplya, A.~Zhidenko and A.~F.~Zinhailo,
Class. Quant. Grav. \textbf{36}, 155002 (2019)
doi:10.1088/1361-6382/ab2e25
[arXiv:1904.10333 [gr-qc]].

\bibitem{Konoplya:2011qq}
R.~A.~Konoplya and A.~Zhidenko,
Rev. Mod. Phys. \textbf{83}, 793-836 (2011)
doi:10.1103/RevModPhys.83.793
[arXiv:1102.4014 [gr-qc]].

\bibitem{Konoplya_wkb}R. A. Konoplya, 
{Phys. Rev. D {\bf 68}, 024018 (2003)} [arXiv:gr-qc/0303052].

\bibitem{Maty_wkb}J. Matyjasek and M. Telecka, 
{Phys. Rev. D {\bf 100}, 124006 (2019)} [arXiv:1908.09389].


\bibitem{Cuyubamba:2016cug}
M.~A.~Cuyubamba, R.~A.~Konoplya and A.~Zhidenko,
Phys. Rev. D \textbf{93}, no.10, 104053 (2016)
doi:10.1103/PhysRevD.93.104053
[arXiv:1604.03604 [gr-qc]].

\bibitem{Konoplya:2022kld}
R.~A.~Konoplya and A.~Zhidenko,
JCAP \textbf{11}, 028 (2022)
doi:10.1088/1475-7516/2022/11/028
[arXiv:2210.04314 [gr-qc]].

\bibitem{Zinhailo:2019rwd}
A.~F.~Zinhailo,
Eur. Phys. J. C \textbf{79}, no.11, 912 (2019)
doi:10.1140/epjc/s10052-019-7425-9
[arXiv:1909.12664 [gr-qc]].

\bibitem{Schutz}B. F. Schutz and C. M. Will, 
{The Astrophysical Journal {\bf 291}, L33 (1985)}.

\bibitem{Will_wkb}S. Iyer and C. M. Will, 
{Phys. Rev. D {\bf 35}, 3621 (1987)}.


\bibitem{Gogoi:2023fow}
D.~J.~Gogoi, A.~\"Ovg\"un and D.~Demir,
Phys. Dark Univ. \textbf{42}, 101314 (2023)
doi:10.1016/j.dark.2023.101314
[arXiv:2306.09231 [gr-qc]].

\bibitem{Sekhmani:2023ict}
Y.~Sekhmani and D.~J.~Gogoi,
Int. J. Geom. Meth. Mod. Phys. \textbf{20}, no.09, 2350160 (2023)
doi:10.1142/S0219887823501608
[arXiv:2306.02919 [gr-qc]].
\bibitem{Lambiase:2023hng}
G.~Lambiase, R.~C.~Pantig, D.~J.~Gogoi and A.~\"Ovg\"un,
Eur. Phys. J. C \textbf{83}, no.7, 679 (2023)
doi:10.1140/epjc/s10052-023-11848-6
[arXiv:2304.00183 [gr-qc]].

\bibitem{Gogoi:2023kjt}
D.~J.~Gogoi, A.~\"Ovg\"un and M.~Koussour,
Eur. Phys. J. C \textbf{83}, no.8, 700 (2023)
doi:10.1140/epjc/s10052-023-11881-5
[arXiv:2303.07424 [gr-qc]].

\bibitem{Parbin:2022iwt}
N.~Parbin, D.~J.~Gogoi, J.~Bora and U.~D.~Goswami,
Phys. Dark Univ. \textbf{42}, 101315 (2023)
doi:10.1016/j.dark.2023.101315
[arXiv:2211.02414 [gr-qc]].

\bibitem{Gogoi:2022ove}
D.~J.~Gogoi and U.~D.~Goswami,
JCAP \textbf{02}, 027 (2023)
doi:10.1088/1475-7516/2023/02/027
[arXiv:2208.07055 [gr-qc]].

\bibitem{Hod:2006jw}
S.~Hod,
Phys. Rev. D \textbf{75}, 064013 (2007)
doi:10.1103/PhysRevD.75.064013
[arXiv:gr-qc/0611004 [gr-qc]].

\bibitem{Berti:2003jh}
E.~Berti, V.~Cardoso, K.~D.~Kokkotas and H.~Onozawa,
Phys. Rev. D \textbf{68}, 124018 (2003)
doi:10.1103/PhysRevD.68.124018
[arXiv:hep-th/0307013 [hep-th]].

\bibitem{Ghosh:2005aq}
A.~Ghosh, S.~Shankaranarayanan and S.~Das,
Class. Quant. Grav. \textbf{23}, 1851-1874 (2006)
doi:10.1088/0264-9381/23/6/003
[arXiv:hep-th/0510186 [hep-th]].

\bibitem{Churilova:2019sah}
M.~S.~Churilova and Z.~Stuchlik,
Annals Phys. \textbf{418}, 168181 (2020)
doi:10.1016/j.aop.2020.168181
[arXiv:1910.12660 [gr-qc]].

\bibitem{Malybayev:2021lfq}
A.~N.~Malybayev, K.~A.~Boshkayev and V.~D.~Ivashchuk,
Eur. Phys. J. C \textbf{81}, no.5, 475 (2021)
doi:10.1140/epjc/s10052-021-09252-z
[arXiv:2103.10920 [gr-qc]].

\bibitem{Gogoi:2024vcx}
D.~J.~Gogoi and S.~Ponglertsakul,
[arXiv:2402.06186 [gr-qc]].

\bibitem{Sekhmani:2024fjn}
Y.~Sekhmani, D.~J.~Gogoi, M.~Koussour, R.~Myrzakulov and J.~Rayimbaev,
Phys. Dark Univ. \textbf{44}, 101442 (2024)
doi:10.1016/j.dark.2024.101442

\bibitem{Gogoi:2023ffh}
D.~J.~Gogoi, J.~Bora, M.~Koussour and Y.~Sekhmani,
Annals Phys. \textbf{458}, 169447 (2023)
doi:10.1016/j.aop.2023.169447
[arXiv:2306.14273 [gr-qc]].


\bibitem{Vagnozzi:2022moj}
S.~Vagnozzi, R.~Roy, Y.~D.~Tsai, L.~Visinelli, M.~Afrin, A.~Allahyari, P.~Bambhaniya, D.~Dey, S.~G.~Ghosh and P.~S.~Joshi, \textit{et al.}
Class. Quant. Grav. \textbf{40}, no.16, 165007 (2023)
doi:10.1088/1361-6382/acd97b
[arXiv:2205.07787 [gr-qc]].


\bibitem{EslamPanah:2020hoj}
B.~Eslam Panah, K.~Jafarzade and S.~H.~Hendi,
Nucl. Phys. B \textbf{961}, 115269 (2020)
doi:10.1016/j.nuclphysb.2020.115269
[arXiv:2004.04058 [hep-th]].


\bibitem{Gogoi:2023ntt}
D.~J.~Gogoi, Y.~Sekhmani, D.~Kalita, N.~J.~Gogoi and J.~Bora,
Fortsch. Phys. \textbf{71}, no.4-5, 2300010 (2023)
doi:10.1002/prop.202300010
[arXiv:2306.02881 [gr-qc]].
\bibitem{Amir:2017slq}
M.~Amir, B.~P.~Singh and S.~G.~Ghosh,
Eur. Phys. J. C \textbf{78}, no.5, 399 (2018)
doi:10.1140/epjc/s10052-018-5872-3
[arXiv:1707.09521 [gr-qc]].

\bibitem{Johannsen:2013vgc}
T.~Johannsen,
Astrophys. J. \textbf{777}, 170 (2013)
doi:10.1088/0004-637X/777/2/170
[arXiv:1501.02814 [astro-ph.HE]].



\bibitem{Das:2019sty}
A.~Das, A.~Saha and S.~Gangopadhyay,
Eur. Phys. J. C \textbf{80}, no.3, 180 (2020)
doi:10.1140/epjc/s10052-020-7726-z
[arXiv:1909.01988 [gr-qc]].



\bibitem{Wei2013}S.-W. Wei and Y.-X. Liu, 
J. Cosmol. Astropart. Phys. {\bf 11}, 063 (2013).

\bibitem{emission2}D. N. Page, Phys. Rev. D {\bf 13}, 198 (1976).


\bibitem{Cai:2020igv}
X.~C.~Cai and Y.~G.~Miao,
Eur. Phys. J. C \textbf{81}, no.6, 559 (2021)
doi:10.1140/epjc/s10052-021-09351-x
[arXiv:2011.05542 [gr-qc]].

\bibitem{Decanini:2011xi}
Y.~Decanini, G.~Esposito-Farese and A.~Folacci,
Phys. Rev. D \textbf{83}, 044032 (2011)
doi:10.1103/PhysRevD.83.044032
[arXiv:1101.0781 [gr-qc]].

\bibitem{Cuadros-Melgar:2020kqn}
B.~Cuadros-Melgar, R.~D.~B.~Fontana and J.~de Oliveira,
Phys. Lett. B \textbf{811} (2020), 135966
doi:10.1016/j.physletb.2020.135966
[arXiv:2005.09761 [gr-qc]].




\end{thebibliography}
\end{document}